\documentclass[12pt,twoside]{article}
\pdfoutput=1

\newlength{\dinwidth}
\newlength{\dinmargin}
\usepackage{amsmath, amsthm, amsfonts, amssymb, bm, mathrsfs, indentfirst,
sectsty, graphicx, fancyhdr, slashed, fullpage, color, authblk}

\theoremstyle{definition}

\theoremstyle{remark}

\def\beq{\begin{equation}}
\def\enq{\end{equation}}

\newcommand{\be}{\begin{equation}} \newcommand{\ee}{\end{equation}}
\newcommand{\bea}{\begin{eqnarray}} \newcommand{\eea}{\end{eqnarray}}
\newcommand{\beann}{\begin{eqnarray*}}  \newcommand{\eeann}{\end{eqnarray*}}
\newcommand{\bfig}{\begin{figure}} \newcommand{\efig}{\end{figure}}
\newcommand{\ba}{\begin{array}} \newcommand{\ea}{\end{array}}
\newcommand{\bcen}{\begin{center}} \newcommand{\ecen}{\end{center}}
\newcommand{\btab}{\begin{tabular}} \newcommand{\etab}{\end{tabular}}

\newcommand{\matt}{\left ( \begin{array}{ccc}}
\newcommand{\ematt}{\end{array} \right )} \newcommand{\matf}{\left
(\begin{array}{cccc}}
    \newcommand{\ematf}{\end{array} \right )} \newcommand{\vect}{\left (
\begin{array}{c}}
    \newcommand{\evect}{\end{array} \right )}    \def\beqn{\begin{eqnarray}}
 \def\eeqn{\end{eqnarray}}  
     
\newtheorem{Proposition}{Proposition}[section]

\newtheorem{Theorem}{Theorem}[section]
\newtheorem{Lemma}{Lemma}[section]
\newtheorem{Corrolary}{Corrolary}[section]

\newcommand{\bp}{\begin{Proposition}}   \newcommand{\ep}{\end{Proposition}}
\newcommand{\bt}{\begin{Theorem}}   \newcommand{\et}{\end{Theorem}}
\newcommand{\bl}{\begin{Lemma}}     \newcommand{\el}{\end{Lemma}}
\newcommand{\bc}{\begin{Corrolary}} \newcommand{\ec}{\end{Corrolary}}

\begin{document}

\begin{center}
 {\Large \bf \sc Inhomogeneous superfluids}
\\[1.52cm]
{\large  Ignacio Salazar Landea\footnote{peznacho@gmail.com} }

\bigskip

{\small \it Instituto de F\'\i sica La Plata (IFLP) and Departamento
de F\'\i sica Universidad Nacional de La Plata, CC 67,
1900 La Plata, Argentina}\\
\end{center}
\bigskip

\begin{abstract}
We show examples of a striped superfluid in a simple
$\lambda\varphi^4$ model at finite velocity and chemical potential
with a global $U(1)$ or $U(2)$ symmetry. Whenever the chemical
potential is large enough we find flowing homogeneous solutions and
static inhomogeneous solutions at any arbitrary small velocity. For
the $U(1)$ model the inhomogeneous solutions found are energetically
favorable for large enough superfluid velocity and the homogeneous
and inhomogeneous phases are connected via a first order phase
transitions. On the other hand, the $U(2)$ model becomes striped as
soon as we turn on the velocity through a second order phase
transition. In both models increasing the velocity leads to a second
order phase transition into a phase with no condensate.

\end{abstract}

\tableofcontents

\section{Introduction}
In this paper we will address the existence of inhomogeneous or
striped superfluids. We have a two folded motivation to do so.

Firstly, superfluidity is an important phenomenon in both condensed
matter and high energy physics. While superfluidity was first
measured and studied in cold helium, it's also believed to play a
role in high density states of nuclear matter.

On the other hand, gravitational duals of superfluids have been
proposed recently in \cite{Hartnoll:2008vx} and a recipe for making
them flow was proposed in \cite{Basu:2008st,Herzog:2008he}. A linear
analysis of the stability was made in \cite{Amado:2013aea}, where a
striped instability was found, with similar features to those found
in the weak coupling limit \cite{Alford:2012vn,Alford:2013koa}.
Furthermore, evidence of the existence of such an inhomogeneous
phase was also found in the $T=0$ limit of a similar model in
\cite{Arean:2011gz}. Similar hints were also found in brane models
in \cite{Jokela:2014wsa}.

Back to the real world, superfluid states of matter are likely to
exist in the interior of compact stars. Neutrons living in the
interior of a neutron star as well as quarks inside a hybrid star
may become superfluid trough Cooper pairing. In this context, the
works \cite{Alford:2012vn, Alford:2013koa} have recently shown that
Landau's $\lambda\varphi^4$ model has instabilities apparently
towards an inhomogeneous phase when the superfluid flows fast
enough. A similar effect was found for two Bose-Einstein condensates
described by the Gross-Pitaevskii model for large enough relative
velocity \cite{Yuka} or for a single Bose-Einstein condensate with
finite range interactions \cite{Kuni}.

Also a $U(2)$ Landau's $\lambda\varphi^4$ model was introduced in
the context of Kaon condensation \cite{Bedaque:2001je}. When the
condensation is induced by a strangeness chemical potential,
Goldstone modes with a quadratic in momentum dispersion relation
appear. The existence of such modes suggest that the theory will not
be able to accommodate a superflow.

 In this paper we
will study the zero temperature Landau's $\lambda\varphi^4$ theory
in the presence of superfluid velocity for both the $U(1)$ and
$U(2)$ models. We will define the superfluid four-velocity as
\cite{Hoyos:2014nua}  \bea\label{hoyospmu} \xi_\mu=i g^{-1}
\partial_\mu g + A_\mu\,, \eea where $g$ is an element of the
symmetry group and $A_\mu$ an external gauge field. With this
definition the superfluid velocity will be basically the derivative
of the Goldstone boson in the broken phase.

 For the
$U(1)$ model we find a first order phase transition at a critical
velocity $v_c\approx0.365$ at which the homogeneous solution is no
longer preferred. On the other hand, for the $U(2)$ model we find
that the inhomogeneous solution has a smaller energy for arbitrary
small values of the velocity. In both cases the inhomogeneous
solution is static.

As a final remark, global gauge fields can also be used to account
the geometrical frustration of a system. In this context interesting
phase transitions were found towards phases with modulated
correlations for $S0(n>2)$ fields \cite{Nussinov:2003wg}. Also, the
existence of excitations with minimum at finite momentum, kills the
ordering at finite temperature leading to a generalized version of
Coleman-Mermin-Wagner theorem. Similar results were also obtained in
lattice models in \cite{Nussinov1, Nussinov2, Nussinov3}.

\section{Inhomogeneous $U(1)$ superfluid}

\subsection{The model}

Lets consider a global $U(1)$ Landau's $\lambda\varphi^4$ just like
in \cite{Alford:2012vn}. \bea {\cal
L}=\partial_\mu\varphi\partial^\mu\varphi^*-m^2|\varphi|^2-\lambda
|\varphi|^4\, , \eea where $\varphi$ is a complex scalar field, $m
\geq 0$ its mass and the coupling constant $\lambda>0$. The
lagrangian is invariant under $U(1)$ rotations $\varphi\rightarrow
e^{i \alpha}\varphi$ which implies a conserved current. Notice that
the condition of having a real mass implies that the quadratic term
of the potential will be positive and that the spontaneous symmetry
breaking occurs only if we introduce a chemical potential associated
to the conserved current. If the chemical potential is greater then
the mass $m$ we get a Bose-Einstein condensate. In this formalism we
will introduce a chemical potential trough a temporal dependence in
the phase of the order parameter.

We will choose the following ansatz for the Bose-Einstein condensate
\bea \varphi(x)=\frac{e^{i\psi(\vec{x})}}{\sqrt{2}}\rho(\vec{x})
\,.\eea Here, $\rho(\vec x)$ is the modulus and $\psi(\vec x)$ the
phase of the condensate. Introducing this ansatz in the condensate
we get the following tree level lagrangian \bea {\cal
L}=\frac12\partial_\mu \rho
\partial^\mu \rho+\frac{\rho^2}{2}\left(\partial_\mu
\psi\partial^\mu\psi-m^2\right)-\frac\lambda4 \rho^4 \,.\eea

The classical equations of motion read \bea \label{eomr}
\Box\rho&=&\rho\left(\sigma^2-m^2-\lambda\rho^2\right)\, ,
\\\label{eomp}
\partial_\mu\left(\rho^2\partial^\mu\psi\right)&=&0\, ,
\eea where we have called \bea
\sigma^2\equiv\partial_\mu\psi\partial^\mu\psi \,.\eea

A simple classical solution to the equations of motion is \bea
\label{simplesolp} \psi&=&p_\mu x^\mu\,, \\\label{simplesolr}
\rho&=&\sqrt{\frac{p^2-m^2}{\lambda}} \,.\eea This ansatz
corresponds to an infinite superfluid flowing uniformly. The density
and flow are determined by the components of $p_\mu$, that are
simply numbers, they do not depend on $\vec x$ and they are not
determined by the equations of motion. The value of $p_\mu$ is
determined by the boundary conditions, that specify the topology of
the field configuration, i.e. the winding of the phase of the order
parameter as we cross the space-time region in which the superfluid
lives.

Notice that in this case $p_\mu$ is nothing but the superfluid
four-velocity defined by equation (\ref{hoyospmu}). We will
consider, without any loss of generality, our four-velocity to have
nontrivial components only in the $t$ and $x$ directions resulting
on \bea p_\mu=(\mu,v,0,\dots)\,. \eea Here $\mu$ is the chemical and
$v$ the superfluid velocity.

From now on we will consider the $m=0$ case where we can scale away
$\lambda$. Furthermore we can measure the velocity $v$ in terms of
the chemical potential, or equivalently we will set $\mu=1$.

\subsection{Linear analysis}

In this section we shall review the arguments of
\cite{Alford:2012vn,Alford:2013koa} about the existence of a
instability towards an inhomogeneous phase. We will always work in
the zero temperature limit.

Lets consider the following ansatz \bea
\varphi=(\rho+\delta\rho)e^{i\psi+i\delta\alpha}\,, \eea where
$\rho$ and $\psi$ are given by the classical homogeneous solutions
(\ref{simplesolp}-\ref{simplesolr}) and $\delta\alpha$ and
$\delta\rho$ are small fluctuations around that solution. The
linearized equations of motion read \bea \Box
\delta\rho-2(1-v^2)\delta\rho+2\sqrt{1-v^2}\left(\partial_t\delta\alpha-v\partial_x\delta\alpha\right)&=&0\,,\\
\Box\delta\alpha+\frac2{\sqrt{1-v^2}}\left(\partial_t\delta\rho-v\partial_x\delta\rho\right)&=&0\,,
\eea and we will suppose an harmonic dependence $\approx e^{-i\omega
t+ik_x x+i k_\bot y}$.

We wish to look now for the velocity of the Goldstone modes around
this background. The same can be computed taking simultaneously the
limit of $\omega$, $k_x$ and $k_\bot$ to zero in the mass matrix
\bea \left(
  \begin{array}{cc}
    2(v^2-1)+O(k^2) & -2 i \sqrt{1-v^2}(k_x v+\omega)+O(k^3) \\
   2 i \sqrt{1-v^2}(k_x v+\omega)+O(k^3) & (v^2-1)\left(k_x^2+k_\bot^2-\omega^2+O(k^3)\right) \\
  \end{array}
\right)\,, \eea with $k^2=k_x^2+k_\bot^2$ and
$\tan\theta=k_x/k_\bot$. The dispersion relations are obtained
asking for $\omega$ such that the determinant of the mass matrix is
zero. Then the sound velocity is just the linear in $\vec k$
coefficient of the dispersion relation $\omega\approx \vec{v}_s\cdot
\vec{k}+\dots$. Measured at an angle $\theta$ with respect to the
superfluid velocity it reads \bea \label{soundvel}
v_s=\frac{2v\cos\theta-\sqrt{1-v^2}\sqrt{3-2v^2-v^2\cos2\theta}}{v^2-3}\,.\eea

\begin{figure}[htp!]
\centering
\includegraphics[width=220pt]{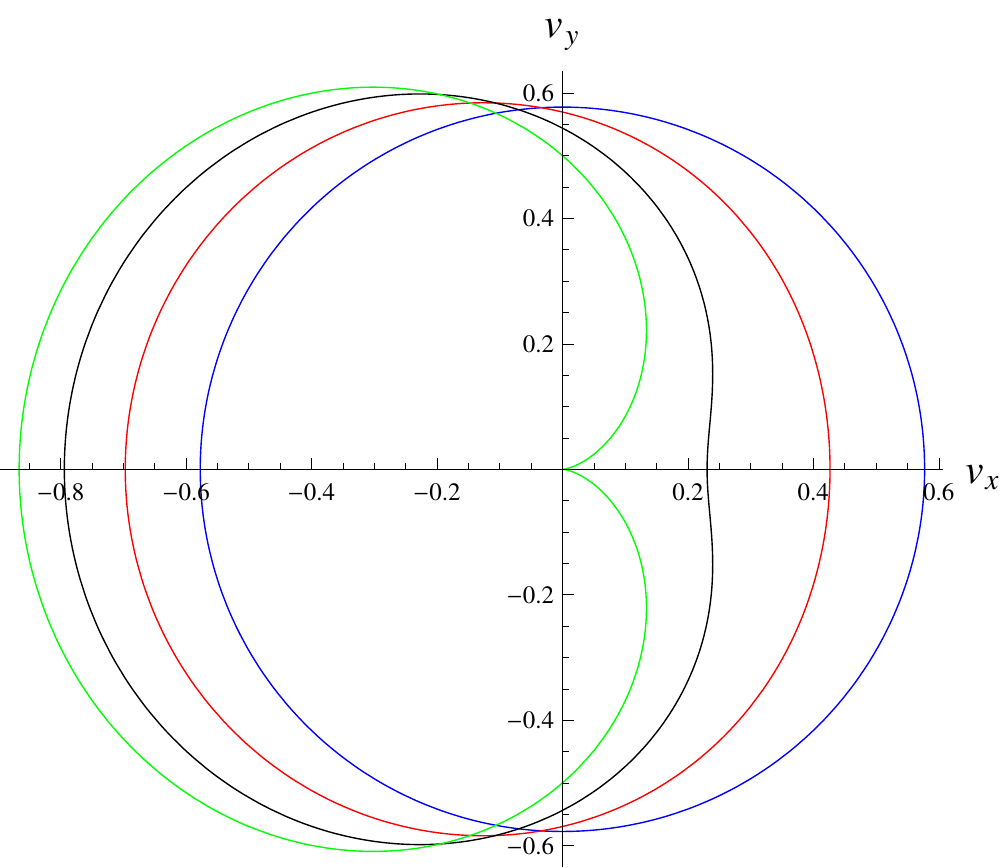} \hfill
\includegraphics[width=220pt]{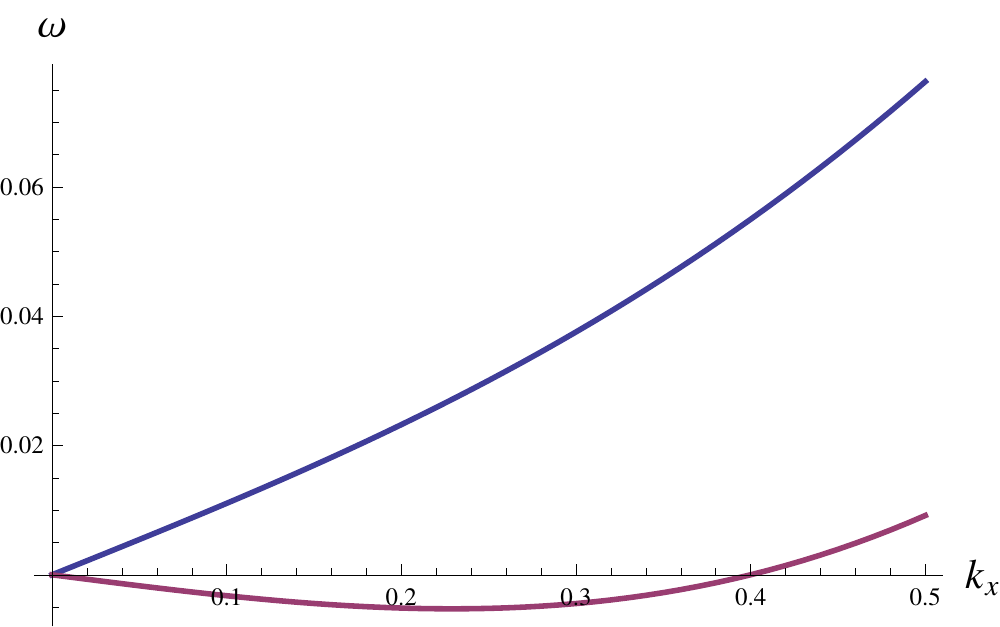}\caption{(Left) Sound velocity for $v=0$ (blue), $v=0.4$ (red), $v=0.6$ (black), $v=1/\sqrt{3}$ (green).
(Right) Dispersion relation beyond the hydrodynamic limit for
$v=0.5$ (blue) and $v=0.6$ (purple) and momentum in the direction of
the superflow. The energy of the purple curve shows a minimum at
finite momentum.}
 \label{soundv}
\end{figure}

On the left hand side of Figure \ref{soundv} we show a parametric
plot of the sound velocity (\ref{soundvel}). As we can see, there is
a critical superfluid velocity $v^*=1/\sqrt{3}$ at which the
excitations have zero velocity. According to the Landau criterion
this might signal an instability which shall kill the superflow.

Going beyond the small momentum limit might give some insight on the
fate of the superfluid after the instability is triggered. Indeed,
as showed on the right hand side of Figure \ref{soundv}, for
superflow velocities greater than $v^*$ the energy has a minimum at
finite momentum, which signals an instability towards an
inhomogeneous phase.

We have showed evidence for the existence of a phase transition to a
non homogeneous phase at large enough superfluid velocities. In the
next section we will address the issue of the construction of such a
phase. The construction itself will show that the second order phase
transition argued in this section is actually over-seeded by a first
order phase transition at a smaller velocity.

\subsection{Construction of the inhomogeneous phase}

In this section we shall construct the inhomogeneous phase whose
existence was hinted in the previous section. In order to do so we
will add a spatial dependence to the ansatz for the fields that does
not change the boundary conditions, so that the winding of the phase
across the superfluid does not change. A natural way to do so is to
add a periodic $x$ dependence.

 To begin
with, let us consider the following ansatz \bea \rho\equiv\rho(x)\,,\\
\psi\equiv-\mu \,t+\alpha(x)\,. \eea
 Let us expand $\rho(x)$ and
$\alpha'(x)$ alla Fourier \bea \rho(x)=\sum_{n=0}^{n_{max}} \rho_n
\cos(n k x)\,,\\
 \alpha'(x)=\sum_{n=0}^{n_{max}} \alpha_n \cos(n k x)\,, \eea where
$n_{max}$ will be a numerical cutoff and the zeroth coefficient
gives the velocity $\alpha_0\equiv v$.

\begin{figure}[htp!]
\centering
\includegraphics[width=220pt]{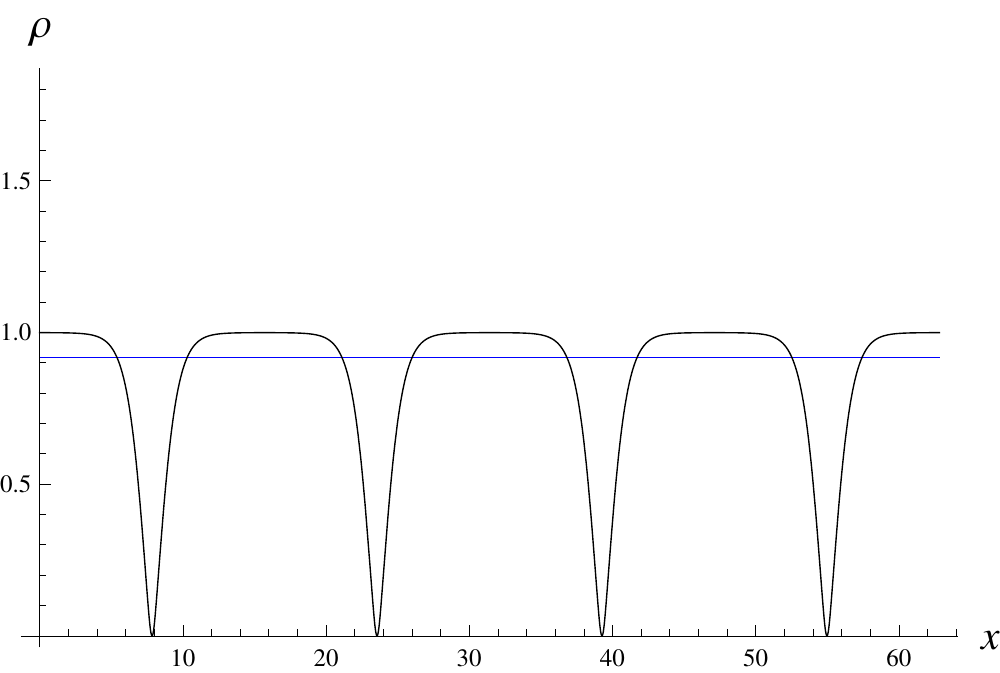} \hfill
\includegraphics[width=220pt]{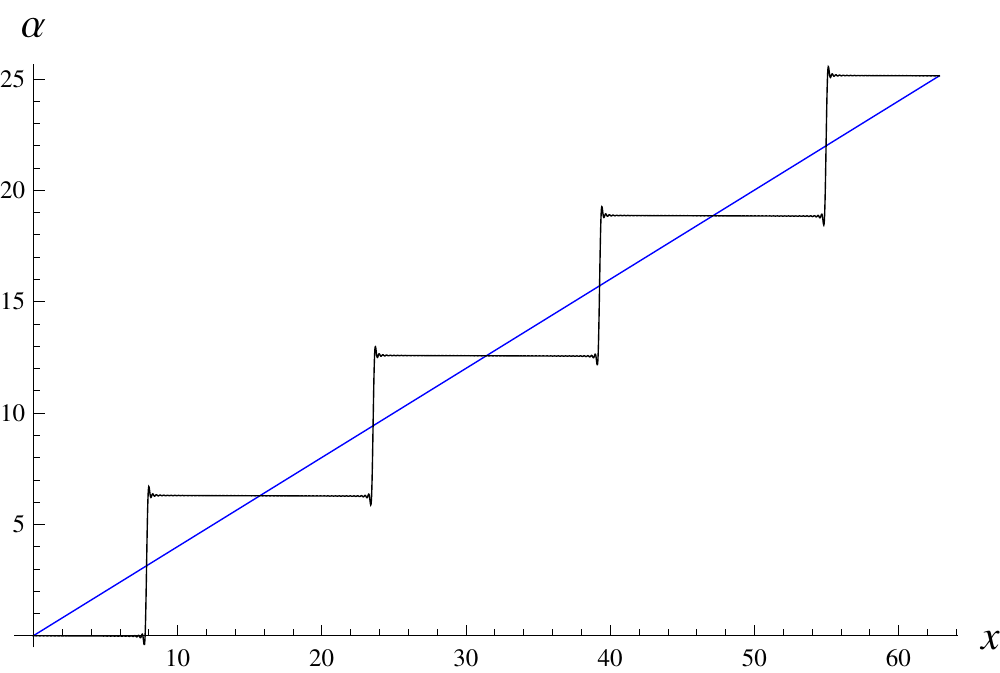}
\caption{Modulus (left) and phase (right) for the homogeneous (blue)
and inhomogeneous (black) solutions. This solutions correspond to
$v=0.4$ and $n_{max}=60$.}
 \label{sols}
\end{figure}

We can now use \emph{Mathematica's FindRoot} command to numerically
integrate the equations of motion (\ref{eomr}-\ref{eomp}). The
procedure consists in solving the equations for each Fourier mode up
to a maximum $n_{max}$, chosen large enough so that we can trust
that the solution won't change significatively if further modes are
taken into account. Two kind of solutions were found numerically. A
first one in which all modes are null but the zero modes, and
corresponds to the homogeneous phase. The second one is spatially
modulated and the phase looks like a step function in the large
$n_{max}$ limit. An example of both solutions can be observed in
Figure \ref{sols}

The picture would be the following. For certain critical velocity
$v_c$ there is a first order phase transition. For low velocities
$v<v_c$ the modulus of the condensate is constant and its phase is
linear in $x$. As we increase the the velocity the phase at which
the condensate is inhomogeneous dominates. In these solutions the
phase of the order parameter is no longer linear with $x$, but a
stairway of $v$ sized steps. Since there is no continuous connection
between homogeneous and inhomogeneous solutions at finite $v$, the
phase transition must be first order. Being that the case, the
linear analysis of the previous section will not shed light on this
phase transition.

\begin{figure}[htp!]
\centering
\includegraphics[width=250pt]{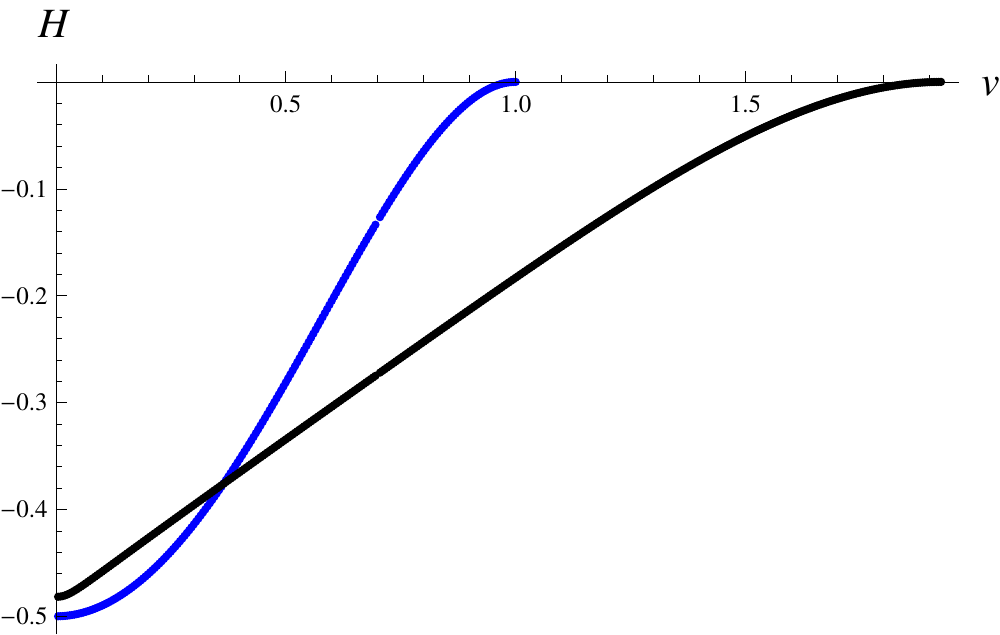}\caption{Energy density of the homogeneous (blue) and inhomogenous (black) solutions.  We
have used $n_{max}=60$ to generate this plot.}
 \label{FE}
\end{figure}

In Figure \ref{FE} we plot the energy density of the homogeneous and
inhomogeneous solutions. As we can see from the plot, the phase
transition is indeed first order since the energy of the system (the
energy of the solution with the lowest energy) has a discontinuity
in its derivative, resulting from the fact that the inhomogeneous
solution does not arise continuously from the homogeneous one.  The
critical velocity, where the energy of the different solutions is
the same, is $v_c\approx0.365<1/\sqrt{3}$. This first order phase
transition overseeds the lineal instability of the homogeneous
background, and the stability analysis should be redone considering
fluctuations around this new background.

Another interesting fact that can be observed in Figure \ref{FE} is
that while the homogeneous solutions with $\rho\neq0$ only exist for
$v<1$, inhomogeneous solutions exist for $v<\tilde v\approx1.92$.
The complete phase diagram would read as follows. For velocities
larger than $\tilde v$ the system lives in the trivial solution,
with no condensate. As we lower the velocity, a second order phase
transition into the inhomogeneous solution occurs at $\tilde v$. If
we keep on lowering the velocity a first order phase transition
occur at $v_c$ into the homogenous phase. This phase dominates down
to null velocity.

The existence of solutions with velocity larger than $v=1$ is very
counterintuitive and may even seem unphysical but are certainly
needed to smoothly connect with the normal phase. This should not be
an issue since the 
phase uses the places in space to jump up to the next step 
precisely where the modulus $\rho$ is zero.
Then we must analyze carefully the superfluid current \bea
j_\mu=\rho(\partial_\mu \alpha+A_\mu)\,, \eea which is a physical
observable. When doing so we reach to the conclusion that even
though the average velocity is grater than one, one can see that the
supercurrent is time-like everywhere. Furthermore, $x$ component
related to the flow is exactly zero everywhere since the phase is
constant everywhere but in the places where the $\rho=0$.  We reach
then to the conclusion that for high enough superfluid velocities
the system goes trough a first order phase transition into a static
striped phase.


\section{Inhomogeneous $U(2)$ superfluid}

\subsection{The model}

Inspired by Kaon condensation in the color-flavor locked phase of
QCD the authors of \cite{Schafer:2001bq, Miransky:2001tw} studied
QCD at finite strangeness chemical potential. It was shown that at a
critical chemical potential equal to the mass of the Kaons, Kaon
condensation occurs through a continuous phase transition.
Furthermore, a Goldstone boson with a non relativistic dispersion
relation  $\omega\sim p^2$ appears in the condensed phase spectrum.
To illustrate such a fact the authors consider the following
 model:
\begin{equation}
  {\cal L} = -(\partial_0 -i \mu)\phi^\dagger
  (\partial_0 +i \mu)\phi + \partial_i\phi^\dagger\partial_i\phi
  + M^2 \phi^\dagger \phi + \frac\lambda2
  (\phi^\dagger \phi)^2\,,
  \label{Lagr}
\end{equation}
where $\phi$ is a complex scalar doublet,
\begin{equation}
   \phi = \vect \phi_1 \\ \phi_2 \evect.
\end{equation}
Here we have introduced the chemical potential $\mu$ through an
external gauge field, minimally coupled to the scalar doublet,
following \cite{Schafer:2001bq, Miransky:2001tw}.

While $\mu<M$ the masses of the four excitations of the theory are
the roots $\omega$ of
\begin{align}
 (\omega \pm \mu)^2=M^2\,.
\end{align}
All of them are doubly degenerated. It is immediate to check that as
soon as $\mu=M$ the $U(2)$ symmetry is broken and a new vacuum must
be chosen:
\begin{equation}
 \phi = \vect 0
  \\ \rho_0 \evect, \qquad {\rm with} \quad \rho_0^2 = \frac{\mu^2 -M^2}\lambda\,.
  \label{choice}
\end{equation}

If we study the fluctuations of $\phi$ around this background one
finds two positive massive modes and two non massive modes with
 dispersion relations:
\begin{align}
\omega_1^2&=\frac{\mu^2-M^2}{3\mu^2-M^2}\,k^2+O(k^4)\,,  \label{eq:typeI}\\
\omega_2^2&=6\mu^2-2M^2+O(k^2)\,,\label{eq:gapped}\\
\omega_3^2&= k^2 - 2 \mu \omega_3\,,\label{eq:E3}\\
\omega_4^2&= k^2 + 2\mu \omega_4\,.\label{eq:E4}
\end{align}
If we focus in the positive roots we see that $\omega_1$ is a normal
Goldstone mode with a linear dispersion relation. The positive root
of (\ref{eq:E3}) is
\begin{equation}
 \omega_3 = \frac{k^2}{2\mu}+O(k^4)\,.
\end{equation}
This is by definition a type II Goldstone mode: it has a nonlinear
dispersion relation proportional to an even power of momentum. Since
the  theory has Lorentz symmetry we also have a negative mode with
quadratic dispersion. This comes from the negative root of
$\omega_3$. Finally $\omega_2$ and $\omega_4$ are massive modes with
\begin{equation}
 \omega_4 = 2\mu + O(k^2)\,.\label{eq:spmode}
\end{equation}
Since the symmetry breaking pattern is $U(2) \rightarrow U(1)$ we
have three spontaneously broken generators but only two massless
modes in the spectrum. This is due to the quadratic dispersion
relation, and satisfies Chadha-Nielsen counting rules
\cite{Nielsen:1975hm}. The role of $\omega_4$ is special since it is
the mode that couples with the type II Goldstone mode in
(\ref{eq:E3}) and (\ref{eq:E4}). There is evidence that its energy
at zero momentum is protected under quantum corrections \cite{
Kapustin:2012cr, Brauner:2006xm, Nicolis:2012vf}. Some recent
related papers are
\cite{Watanabe:2011ec,Watanabe:2012hr,Watanabe:2013iia,Hidaka:2012ym}.

As it has been pointed out in \cite{Brown:1995ta}, the existence of
this type II Goldstone modes should make the system unstable when an
arbitrarily small velocity is turned on. We will address this issue
more deeply.

\subsection{Adding velocity naively}
\label{linearv}

Let us naively add a velocity in the $x$ direction by turning on an
external $A_x=v$ gauge field. Immediately we can see that this
contributes to the condensate as a positive mass term, so the
homogeneous classical solution will read
\begin{equation}
 \phi = \vect 0
  \\ \rho_0 \evect, \qquad {\rm with} \quad \rho_0^2 = \frac{\mu^2 -M^2-v^2}\lambda\,.
  \label{choicev}
\end{equation}

We shall consider again perturbations around this background. We can
see that the lower sector is just that of the $U(1)$ sector studied
previously, and will obviously show the same instabilities.

Now let us see what happens to the lower sector. We can see that the
positive branch of the type II Goldstone now acquires a negative
velocity
\begin{equation}
\label{gmode}
 \omega_3 =-\frac{v k_x}{\mu}+ \frac{-(A_x\,k_x)^2+\mu^2(k_x^2+k_\bot^2)}{2\mu^3}+O(k^4)\,.
\end{equation}
This signals the fact that the energy minimum of the perturbation
won't be at zero momentum and that the system will rather be in a
striped phase.

In order to make a more direct connection with the previous section
we will go to the conformal limit where $M=0$, and we can choose
$\lambda=1$. We will work at fixed chemical potential $\mu=1$, or
equivalently we can say that we measure the velocity in terms of the
chemical potential.

\begin{figure}[htp!]
\centering
\includegraphics[width=250pt]{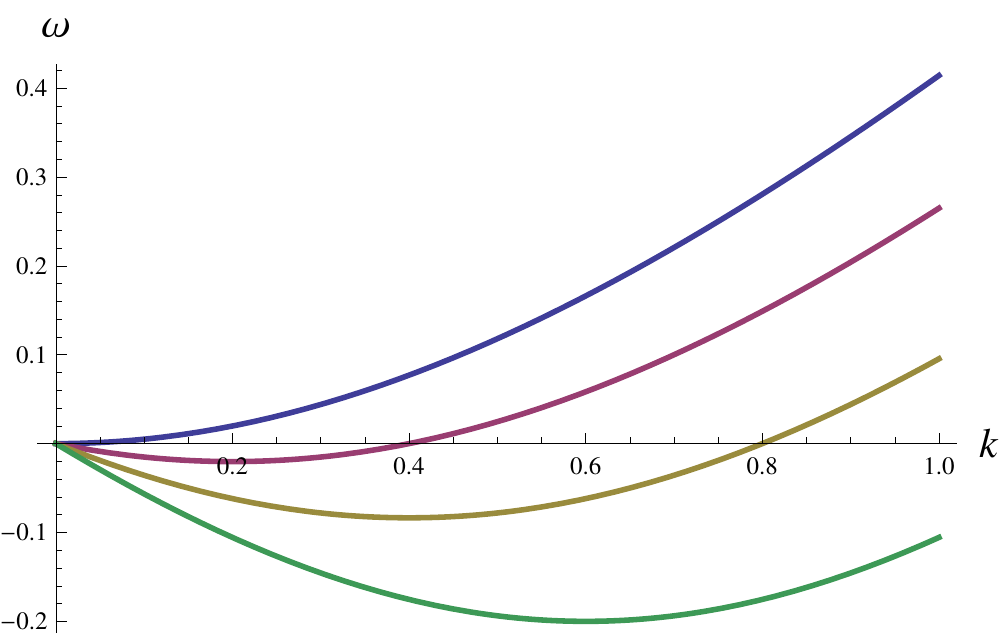}\caption{Dispersion relation of the type II Goldstone mode for $v=0$ (blue), $v=0.2$ (purple),
 $v=0.4$ (yellow) and $v=0.6$ (green).}
 \label{typeII}
\end{figure}

In figure \ref{typeII} we can see the dispersion relation for the ex
type II Goldstone mode $\omega_3$ plotted beyond the hydrodynamic
limit, i.e. considering the expression (\ref{gmode}) at all order in
$k$. As we can see, as we increase the velocity the minimum in
energy occurs at larger momentum. This hints a second order phase
transition to an inhomogeneous phase as soon as we turn on a
velocity.

We will repeat the mechanism of the previous section in order to
construct inhomogeneous solutions corresponding to the $U(2)$
superfluid at finite velocity.

\subsection{Constructing the inhomogeneous phase}

In order to construct the inhomogeneous phase let us consider the
following ansatz
\begin{equation}
 \phi = \vect \rho_u(x) e^{i\alpha_u(x)}
  \\ \rho_d(x) e^{i\alpha_d(x)} \evect\,.
  \label{choicevx}
\end{equation}
The classical equations of motions for the fields read \bea
\rho_u''&=&\left(v+\alpha_u'\right)^2\rho_u+(\rho_d^2+\rho_u^2-1)\rho_u\,,\\
0&=&\rho_u\alpha_u''+2\left(v+\alpha_u'\right)\rho_u'\,,\\
\rho_d''&=&\left(v+\alpha_d'\right)^2\rho_d+(\rho_d^2+\rho_u^2-1)\rho_d\,,\\
0&=&\rho_d\alpha_d''+2\left(v+\alpha_d'\right)\rho_d'\,.
 \eea
Again we will do a Fourier decomposition of the fields \bea
\rho_u(x)=\sum_{n=0}^{n_{max}} \rho_u^{(n)}
\cos(n k x)\,,\\
 \alpha_u'(x)=\sum_{n=1}^{n_{max}} \alpha_u^{(n)} \cos(n k x)\,,\\
\rho_d(x)=\sum_{n=0}^{n_{max}} \rho_d^{(n)}
\cos(n k x)\,,\\
 \alpha_d'(x)=\sum_{n=1}^{n_{max}} \alpha_d^{(n)} \cos(n k x)\,, \eea
where we have removed the zero modes of the phase since they will be
taken into account in the spatial component of the external gauge
field $A_x=v$.

\begin{figure}[htp!]
\centering
\includegraphics[width=220pt]{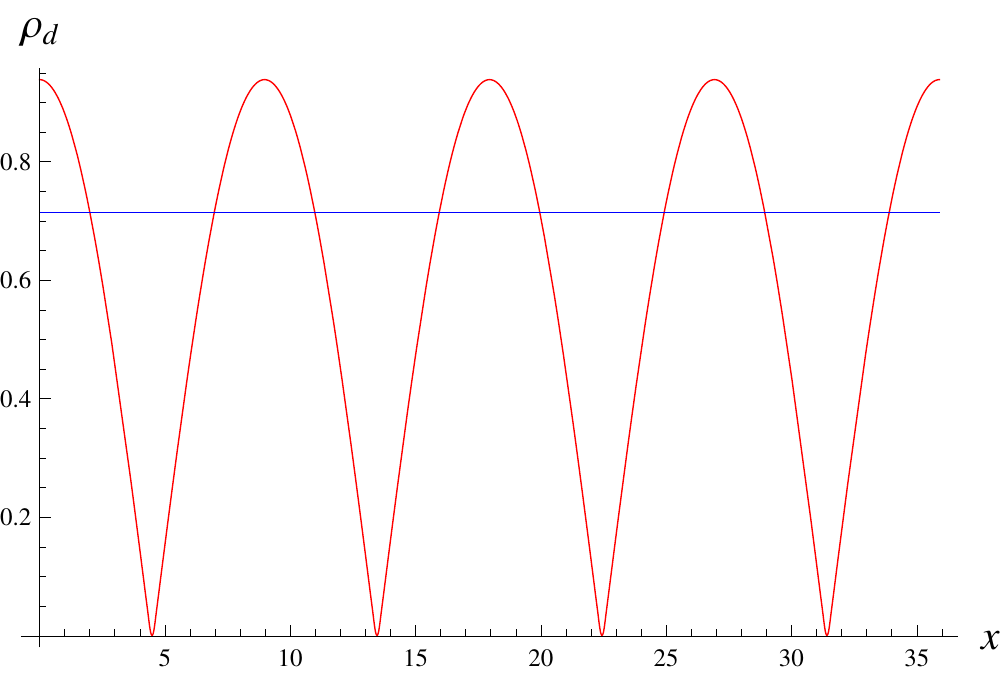} \hfill
\includegraphics[width=220pt]{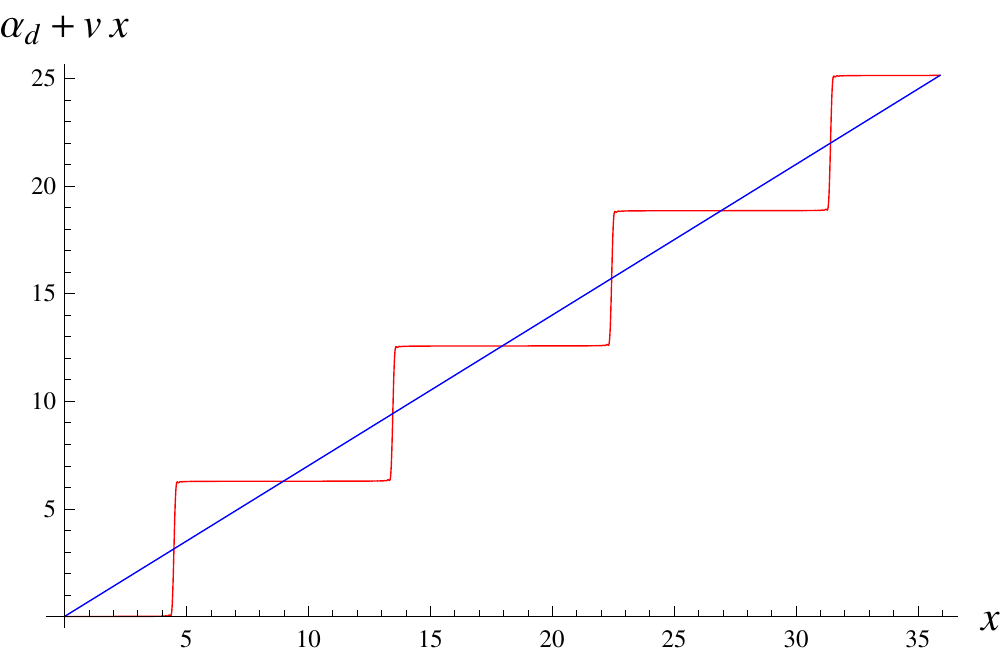}
\includegraphics[width=220pt]{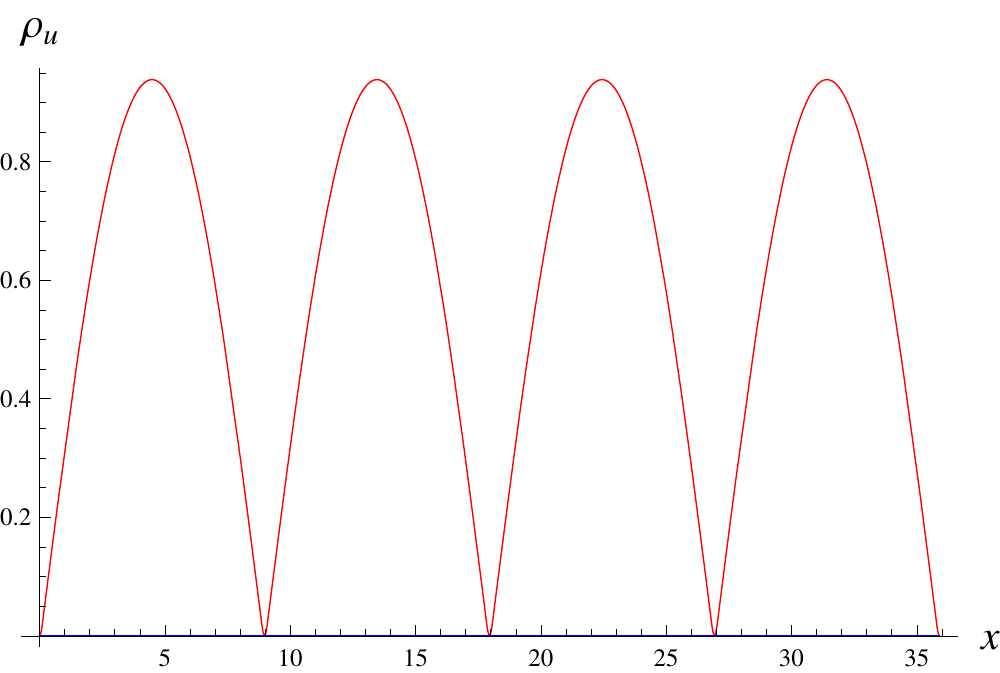} \hfill
\includegraphics[width=220pt]{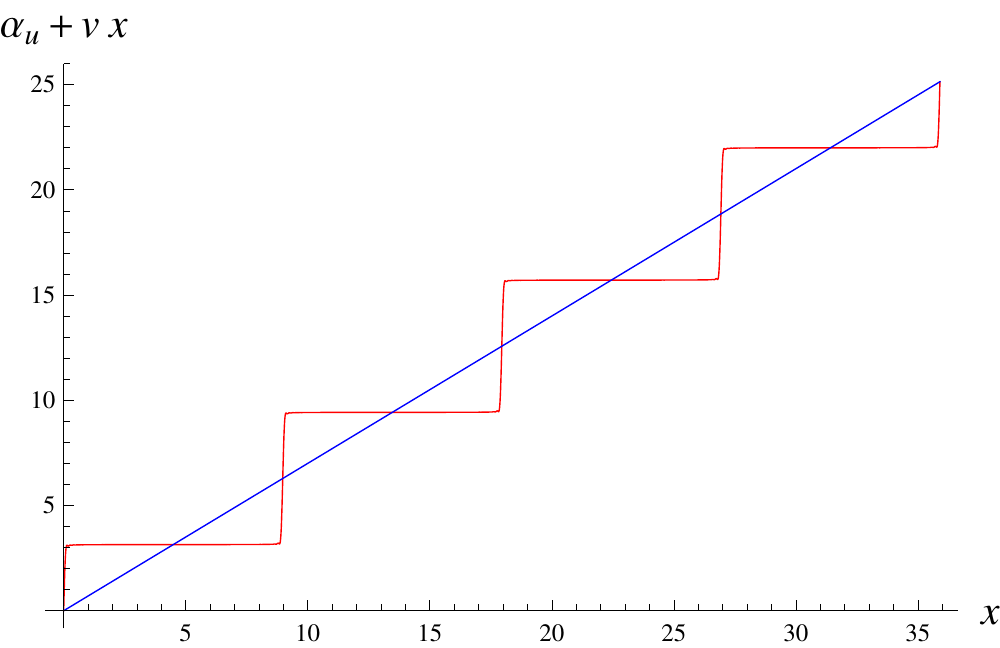}
\caption{Homogeneous (blue) and inhomogeneous (red) solutions for
$v=0.7$ and $n_{max}=60$.}
 \label{sols2}
\end{figure}

We will now solve the equations for each Fourier mode numerically.
Considering $\rho_u=0$ we recover the same solutions that in the
previous section. When we allow a non-trivial profile for $\rho_u$
we find a further numerical solution. Its Fourier coefficients
satisfy \bea \rho_u^{(n)}=(-1)^n\rho_d^{(n)},  \quad
\alpha_u^{(n)}=(-1)^n\alpha_d^{(n)}\,,\eea and correspond to a
solution where both condensates are modulated, with a half period
relative phase in their oscillatory space dependence. Their phases
$\alpha_{u,d}$ are again step functions, and also have a half period
relative phase. An example of these new solutions can be observed in
Figure \ref{sols2}.

\begin{figure}[htp!]
\centering
\includegraphics[width=250pt]{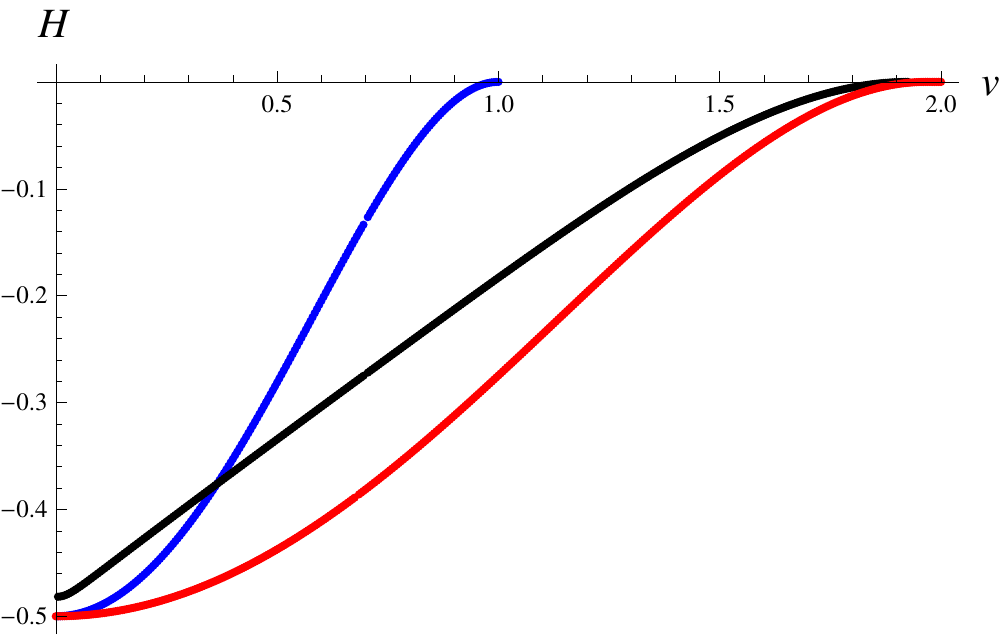}\caption{Energy density of the homogeneous (blue)
and inhomogenous solutions with condensate only in the lower
component (black) and with condensate in both components (red). We
have used $n_{max}=40$ to generate this plot.}
 \label{FEU2}
\end{figure}

In Figure \ref{FEU2} we show the energy density of this new
configuration in contrast to the energy density of the solutions
that also exist in the $U(1)$ model, i.e. the homogeneous solution
and the solution with condensate in only one component. We can see
that as soon as we turn on a velocity, a second order phase
transition occurs into an inhomogeneous phase with two spatially
modulated condensates. This is in agreement with the linear analysis
done in Section \ref{linearv}. The order of the phase transition
comes from the fact that the energy has the same slope for both
solutions with respect to the velocity, which might be a bitt
counter intuitive, since one solution does not emerge continuously
from the other.

We can see from Figure \ref{FEU2} that the solution with two
condensates always have the smallest free energy until it no longer
exist at $\tilde v \approx 1.965$. At such critical velocity the
superfluid solution is connected with the trivial solution.

Once again we can numerically check that the supercurrent is zero
everywhere, so the system does not flow even though we turn on a
superfluid velocity, but it develops stripes.


\section{Conclusions}

We have shown an example of a striped superfluid in a simple
$\lambda\varphi^4$ model at finite velocity and chemical potential.
We have studied two models one with global $U(1)$ gauge symmetry and
the other one with $U(2)$.

For the $U(1)$ model the inhomogeneous solutions found are
energetically favorable for large enough superfluid velocity. The
homogeneous and inhomogeneous phases are connected via a first order
phase transition. Increasing the velocity leads to a second order
phase transition into a phase with no condensate. This work somehow
completes the picture shown in \cite{Alford:2012vn,Alford:2013koa},
about this very same model.

For the $U(2)$ model on the other hand, as soon as we turn on the
velocity we end in a striped phase. This is in agreement with Landau
criterion for superfluidity, since this model has zero velocity
excitations. Increasing the velocity leads to a second order phase
transition into a phase with no condensate.

As a possible continuation of this work, we would like to compute a
similar computation in the context of AdS/CFT, following
\cite{Amado:2013aea}. There, it is shown that the holographic $U(2)$
superfluid constructed in \cite{Amado:2013xya, Krikun:2012yj} is
unstable at all the range of velocities numerically reachable, while
the $U(1)$ model of \cite{Herzog:2008he} is only unstable for large
enough velocities, in agreement with the field theoretical
predictions. The explicit construction of the holographic phases
should be an interesting challenge.

Another interesting problem would be to analyze fluctuations around
this inhomogeneous background, in order to address the problem of
stability.

It would also be interesting to make a connection with fluids
physics, looking at the hydrodynamic limit of this theory, following
the steps of \cite{Alford:2012vn}.

\section*{Acknowledgements}
We would like to thank Irene Amado,  Daniel Are\'an, Raul Arias,
God, Carlos Hoyos, Amadeo Jim\'enez, Karl Landsteiner, Luis Melgar,
Andreas Schmitt and MVC for correspondence, inspiration and advise.
Also we would like to special thank to Daniel Are\'an for
``carefully'' reading this manuscript.

\end{document}